\magnification 1200

\font\titlefont=cmss17 at 29.88 true pt
\font\authorfont=cmssi17 at 20.74 true pt
\font\bigaddressfont=cmss12 at 14.4 true pt 
 at 14.4 true pt
 at 27.4 true pt
 at 20.4 true pt
 at 24.4 true pt
\def \mysubmit {}
\def \mypresent {}
\def \docnum #1 { \def \mydocnum {#1}} 
\def \date #1 { \def \mydate {#1}} 
\def \title #1 {\def \mytitle {#1}}
\def \author #1 {\def \myauthor {#1}}
\def \abstract #1 {\def \myabstract {#1}}

\def \tobesubmittedto #1 { \def \mysubmit {\leftskip=0pt plus 1fill \rightskip=0pt plus 1fill 
\hbox{\vbox{\noindent \hfill \it To be 
submitted To #1 \hfill \ }}}}
\def \submittedto #1 { \def \mysubmit {\leftskip=0pt plus 1fill \rightskip=0pt plus 1fill 
\hbox{\vbox{\noindent \hfill \it Submitted 
To #1 \hfill \ }}}}
\def \presentedat #1 { \def \mypresent {\leftskip=0pt plus 1fill \rightskip=0pt plus 1fill 
\hbox{\vbox{\noindent \it Presented 
at #1  }}}}

\def\maketitle{
\let\footnotesize\small
\let\footnoterule\relax

\ifx\mydate\undefined \def \mydate {
\ifcase\month\or
January\or February\or March\or April\or May\or June\or
July\or August\or September\or October\or November\or December\fi
\space\number\day, \number\year} \fi

\ifx\thispagestyle\undefined \nopagenumbers \fi
\ifx\nopagenumbers\undefined {\thispagestyle{empty}} 
      \setcounter{page}{0}%
      \fi
\null
\vskip 20 pt
\rightline{\logo}
\vskip 2 pt
\rightline{\mydocnum}
\rightline{\mydate}
\vskip 20 pt
{\def\\{\break} 
\leftskip=0pt plus 1fill
\rightskip = 0 pt plus 1fill
\parindent 0 pt
\baselineskip 30 pt
\titlefont \hfil \vbox { \mytitle}\hfil }
\vskip 25 pt
{\def \and {\qquad} 
\leftskip=0pt plus 1fill
\rightskip = 0 pt plus 1fill
\parindent 0 pt 
\authorfont
\lineskip 12 pt
\myauthor
\parfillskip=0pt\par
}%
\vskip 20 pt

{\bigaddressfont
\centerline{ Department of Physics}
\centerline{ Manchester University}
\centerline{ England}
}

\ifx\myabstract\undefined {}
\else
\null\vfil\vskip 10 pt
\centerline{ \bf Abstract}
\vskip 10 pt
\myabstract
\fi

\ifx\footline\undefined   
\begin{figure}[b]
\mypresent
\mysubmit
\end{figure}
\mythanks
\setcounter{footnote}{0} 
\vfil
\null
\else                       
\footline={{\baselineskip=10 pt \vbox{\hbox to \hsize {\mypresent} \vskip 5 pt \hbox to \hsize{\mysubmit}}}}
\vfil
\eject
\pageno=1 
\footline={\hss\tenrm\folio\hss}
\fi
				      
\let\thanks\relax
\gdef\mysubmit{}\gdef\present{}
\gdef\mythanks{}\gdef\myauthor{}\gdef\@title{}\let\maketitle\relax}

\def \phonenumber{4178}
\def\today{\number\day/\number\month/\number\year\space\number\hour%
:\number\minute\space\jobname}

\def \logo {\vbox to 23.5 mm {
\hbox {}
\hbox to 47 mm 
{
\hfil} \vfil }}

\def\header #1\par{ \centerline{\it #1}\par}
\def \letterhead {
\voffset -10 mm
{\advance \hsize by 2 cm
\font\address=cmss12
\vskip -9.5 pt
{\address
\vbox {
\vskip 2 mm
\hbox to 6 cm {\hskip -1 cm Department of Physics and Astronomy\hfill }
\hbox to 6 cm {\hskip -1 cm The University of Manchester \hfill }
\hbox to 6 cm {\hskip -1 cm Manchester \hfill }
\hbox to 6 cm {\hskip -1 cm M13 9PL \hfill }
\hbox to 8 cm {\hskip -1 cm Tel 0161-275-\phonenumber\hskip 1.0 cm 
Fax 0161-273-5867\hskip 1.0 cm}
\hfill
\vbox{
\hbox to 6 cm{\includegraphics{/home/roger/tex/ulogo.ps}} 
\vskip 0.5 mm
}
}
}}
}

\def \letter #1 {
\topskip 0 pt
\vsize 599 pt
\nopagenumbers
\letterhead
\vskip 1 mm

\vbox to 3.4 cm {\vfill #1 \vfill}
\vskip 5 mm

\hbox to 2 cm {\hskip -1 cm \leaders\hrule height .5 pt \hfill \hskip 1 cm}

\rightline {\number\day \
\ifcase\month\or January\or February\or March\or April\or
May\or June\or July\or August\or September\or October\or
November\or December\fi \ \number\year}

\vskip 5 mm
}

\def \AddressPPARC #1 { \leftline{#1 }
\leftline{PPARC,}\leftline{Polaris House,}
\leftline{North Star Avenue,}\leftline{SWINDON,}\leftline{SN2 1SZ}}

\def \AddressRAL #1 { \leftline{#1 }
\leftline{HEP Division,}\leftline{Rutherford Appleton Laboratory,}
\leftline{Chilton,}\leftline{Didcot,}\leftline{Oxon}}

\def \AddressRegistrar  #1 {
\leftline{#1 }
\leftline{The Registrar's Department}
\leftline{Main Building}
\leftline{University of Manchester}
\leftline{Oxford Road}
\leftline{Manchester M13 9PL}}

\def \AddressFaculty #1 {
\leftline{#1 }
\leftline{The Faculty of Science}
\leftline{Roscoe Building}
\leftline{University of Manchester}
\leftline{Oxford Road}
\leftline{Manchester M13 9PL}}

\def \AddressPhysics #1 {
\leftline{#1 }
\leftline{Department of Physics}
\leftline{University of Manchester}
\leftline{Oxford Road}
\leftline{Manchester M13 9PL}}

\def \AddressCERN #1/#2 {
\leftline {#1}
\leftline {#2 Division,}
\leftline {CERN,}
\leftline {CH1211 Gen\`eve 23,}
\leftline {Switzerland}}

\def \NIM #1 {{\it Nucl. Instr \& Meth. \/}{\bf A#1}\ }
\def \ZPC #1 {{\it Zeit. Phys. \/}{\bf C#1}\ }
\def \NPB #1 {{\it Nucl. Phys. \/}{\bf B#1}\ }
\def \PLB #1 {{\it Phys. Lett. \/}{\bf B#1}\ }
\def \PL #1 {{\it Phys. Lett. \/}{\bf #1}\ }
\def \PRD #1 {{\it Phys. Rev. \/}{\bf D#1}\ }
\def \PRL #1 {{\it Phys. Rev. Lett.\/}{\bf #1}\ }
\def \PR #1 {{\it Phys. Rev. \/}{\bf #1}\ }
\def \CPC #1 {{\it Comp. Phys. Comm. \/}{\bf #1}\ }

\newcount \eqnumber
\newcount \fignumber
\newcount \highref

\def \eq #1{\global\advance \eqnumber by 1
\let \rrr=\eqnumber
\xdef #1{\the\rrr}
\eqno (\the\eqnumber)
}
\def \fig #1{Figure \global\advance \fignumber by 1 \the\fignumber
\let \rrr=\fignumber
\xdef #1{Figure \the\sss}
}

\newcount\refcheck
\newcount\thisrf
\def\references #1 {
\thisrf=0
\ifnum\refcheck=0
\else
\leftline{\bf References}
\fi
\input #1
\refcheck=1
}

\def\refiopstyle #1:#2;#3\par{
\relax
\ifnum\refcheck=0
\edef \rrr{#2}
\let #1=\rrr
\else
#2 
\ 
#3
\par
\fi
\relax
}

\def\refseq#1{\ifnum#1>\the\highref\global\advance\highref 1  
\ifnum#1>\the\highref
\message{Reference out of Sequence: expecting \the\highref got #1}
\highref=#1\fi\fi}

\def\reference #1:#2\par{\advance \thisrf by 1
\relax
\ifnum\refcheck=0
\let \sss=\thisrf
\edef \rrr{\the\thisrf\noexpand\refseq{\the\thisrf}}
\let #1=\rrr
\else
\the\thisrf
:\ 
#2
\par
\fi
\relax
}

\font\small=cmr8

\def \tickbox {\lower 2 mm \hbox{
\vbox  {\vskip .5 mm\hrule \hbox to 4 mm
{\vrule \strut  \hfill  \vrule}\vfill\hrule} }}

\newcount\secnum
\newcount\examplenum
\newcount\subnum
\newcount\subsubnum
\newcount\chapnum
\font \splash=cmssi17
\font \titlefont=cmss17 scaled \magstep2

\font \address=cmss12
\font \cf=cmbxsl10
\examplenum=0
\secnum=0
\subnum=0

\def\chapter #1 {
\advance\chapnum by 1 \secnum=0 \subnum=0 \subsubnum=0
\vfill
\hbox{\bf \quad Chapter \number \chapnum : #1}}
\def\lecture #1 #2{
\secnum=0 \subnum=0
\hbox{\centerline{\splash SLUO Lecture #1: #2}}
\headline={\ifnum\pageno>1 SLUO Lecture #1 \dotfill #2 \fi}
\footline={\ifnum\pageno>1 \hss --\  \folio \ -- \hss  \else  \fi}
}

\def\subsection#1 \par{\par \advance\subnum by 1
\subsubnum=0
\goodbreak \vskip 0.3cm\leftline{\cf \number \secnum .\number 
\subnum \ #1} \par}

\def\subsubsection#1 \par{\par \advance\subsubnum by 1
\goodbreak \vskip 0.3cm\leftline{\sl \number \secnum .\number 
\subnum .\number \subsubnum \ #1} \par}

\def\section#1\par{\goodbreak \par \advance\secnum by 1 \subnum=0
\vskip 0.5cm\leftline{\bf \number \secnum . \ #1}\vskip 0.02cm \par
\message{ Section  \number \secnum    #1}
}

\long\def\example #1 {\par \advance\examplenum by 1
\vskip 12 pt \goodbreak \boxit {{\bf Example \number\examplenum :} #1 }
}

\long\def\boxit #1{\vbox {\kern-5pt\hrule\hbox{\vrule\kern3pt
           \vbox{\kern3pt #1 \kern3pt} \kern3pt \vrule} \hrule}}

\def\bull#1\par {\item {$\bullet$} #1 \par}

\def\half {\hbox{${1 \over 2}$}}


\title {Asymmetric Statistical Errors}
\docnum {MAN/HEP/04/02}
\date{24/6/2004}
\author {Roger Barlow}
\abstract {
Asymmetric statistical errors arise for experimental results 
obtained by Maximum Likelihood estimation,
in cases where the number of results is finite
and the
log likelihood function is not a symmetric parabola.
This note discusses how 
separate asymmetric errors on a single result 
should be combined,
and how several
results with asymmetric errors should be combined to give an overall 
measurement.
In the process it considers several methods for
parametrising curves that are approximately parabolic.
 
}
\def \sm{\sigma_-}
\def \sp{\sigma_+}
\def \AE #1 #2 #3 {#1_{-#2}^{+#3}}

\maketitle

\section Introduction

When  an experimental result is presented as
 $x^{+\sigma^+}_{-\sigma^-}$ this signifies,  
just as with the
usual form $x\pm\sigma$, that $x$ is the value given by a
 `best' estimate (i.e.
one with good properties of consistency, efficiency, and lack of bias) and
that the 
68\% central confidence region is $[x-\sigma^-,x+\sigma^+]$.

Such asymmetric errors arise through two common causes.
The first is when a nuisance parameter $a$ has a conventional symmetric (even 
Gaussian)
 probability 
distribution, but produces a non-linear effect on the
desired result $x$.  These errors are generally systematic rather than statistical, and their 
probability distribution is generally best considered from a Bayesian
viewpoint. Their treatment has been 
considered in a previous 
note [1].

The second cause of asymmetry is the
extraction of a result $x$ through the maximisation of a likelihood 
function $L(x)$ which is not a symmetric parabola.  This occurs because the
function is in general only parabolic in the limit when the number of 
results $N$, the
number of terms contributing to the sum which makes up the log likelihood,
is large, and for many results this is not the case.  For such a function
the errors are conventionally read off the points at which the log likelihood falls by $\half$ from its peak, though this 
is not exact [2] and it may be better to obtain the errors from a toy Monte Carlo 
computation.

Although such asymmmetric errors are frequently used in the reporting of 
particle physics results, constructive analyses of their use are scarce 
in the literature
[3].

\section Two Combination Problems

The two most significant questions on the manipulation of asymmetric errors
are the 
{\it Combination of Results} and the 
{\it Combination of Errors}. 

\subsection Combination of Results

The first occurs when one has two results
${x_1}^{+\sigma_1^+}_{-\sigma_1^-}$ and
${x_2}^{+\sigma_2^+}_{-\sigma_2^-}$ 
 of the same quantity. 
This arises when two different experiments measure the same quantity.
Assuming that they are compatible (according to some criterion),
one wants the ppropriate value (and errors)
that combines the two.
This is the equivalent of the well-known expression for symmetric errors 
$${x_1/\sigma_1^2 + x_2/\sigma_2^2 \over 
1/\sigma_1^2 + 1/\sigma_2^2 } \pm 
\sqrt{1 \over 1/\sigma_1^2 + 1/\sigma_2^2}\eqno(1)$$ 
If the log likelihood functions $L_1(x_1)$ and $L_2(x_2)$ are known, then
the combined log likelihood is just the sum of the two. The maximum 
can then be 
found and the errors read off the $\Delta ln L =-\half$ points 

The question naturally extends to more than two results, and it is
clearly a desirably property that the operation be associative: if results
are combined pairwise till only one remains, then the pairing strategy
should not effect the result. For the addition of likelihoods this  
obviously holds.

\subsection Combination of Errors

The second question arises when a particular result (taken, without loss 
of generality, as zero)
 is subject to several
separate (asymmetric)
uncertainties, and one needs to quote the overall  uncertainty.
An obvious example would be the uncertainty due to
background subtraction where the background has several different components,
each with asymmetric uncertainties.
This is the equivalent of the well-known expression for symmetric errors
$$\hbox{\rm If } x=x_1+x_2 \qquad \hbox{\rm then} \qquad \sigma^2 = \sigma_1^2 + \sigma_2^2\eqno(2)$$
Again, it is desirable that the operation be associative. 

If the likelihood functions are known then the joint function
$L_1(x_1)L_2(x_2)$ is defined on the $(x_1,x_2)$ plane with its peak 
at (0,0). The uncertainty on the sum $x_1+x_2$
is found by the profiling technique: we find $\hat L(x_1+x_2)$, the
peak value of the likelihood anywhere on the line $x_1+x_2=constant$, and
the $\Delta log L=-\half$ errors can be read off from this [4].

To explain why this works (and when it doesn't), 
consider first a case where the answer is 
easily found: suppose $x_1$ and $x_2$ are both Gaussian, with the same mean $\sigma$.  
The log likelihood can then be rewritten using $u=x_1+x_2$ and $v=x_1-x_2$:
$$-{x_1^2 \over 2 \sigma^2} -{x_2^2 \over 2 \sigma^2} = 
-{(x_1+x_2)^2 \over 4 \sigma^2} - {(x_1-x_2)^2 \over 4 \sigma^2} =
-{u^2 \over 4 \sigma^2} -{v^2 \over 4 \sigma^2} \eqno(3)$$ 
The likelihood is the product of two Gaussians (of width $\sqrt 2 \sigma$),
one in the combination of interest $u$, the other in the
ignorable combination $v$.

Now for some fixed value of $v$, the likelihood for $u$
is a Gaussian of mean zero, and the 68\% 
central confidence region
for $u$ is given by its standard deviation and 
is of half-width $\sqrt 2 \sigma$. 
If $v$ is fixed at some other value, the likelihood for $u$, and the 
deductions that can be drawn from it, are the same,
Thus one can say  `There is a 68\% probability that 
$u$ lies in the region [$-\sqrt 2 \sigma, \sqrt 2 \sigma$],
whatever value of $v$ is chosen', and this
can legitimately be shortened by striking out the final condition.
And the problem is solved.

To apply this technique in some less transparent
 case we need to factorise the likelihood into the form
$L_1(x_1)L_2(x_2)=L_u(u)L_v(v)$
where we have freedom to choose the functions $L_u$, $L_v$, and the 
form $v(x_1,x_2)$.
In some instances this is clearly possible: a double Gaussian with 
$\sigma_1 \neq \sigma_2$ can be factorised using $v=\sigma_2 x_1 - \sigma_1 x_2$.
There are also instances, such as a volcano-crater shaped function,
which are manifestly impossible to factorise.
These
can readily be proposed as counterexamples, but appear somewhat contrived
and it is reasonable to hope that they might not occur in
practical experience, except for very small $N$.

On the grounds that if this factorisation is impossible we can get nowhere,
let us assume it to be true and see where that leads us.
Finding 
the explicit forms of $v$ and $L_v$ is complicated and one would like to avoid
it. This can be done by noting that:

1: For fixed $v$ the shape of the total likelihood as a function of $u$ is the same

2: For fixed $u$ the shape of the total likelihood as a function of $v$ is the same

(1) tells us that we can study the properties of $L_u(u)$ by fixing on
any value of $v$.  (2) tells us that we can fix the value of $v$ by
finding the maximum,  the likelihood (as a function of $v$, with $u$ fixed)  will
always peak at the same value of $v$.  Thus for a given 
$u=x_1+x_2$
one finds the value of $x_1-x_2$ at which $L$ is greatest, as that is
always the same value of $v$. 

\vbox{
\vskip 4 cm
\includegraphics{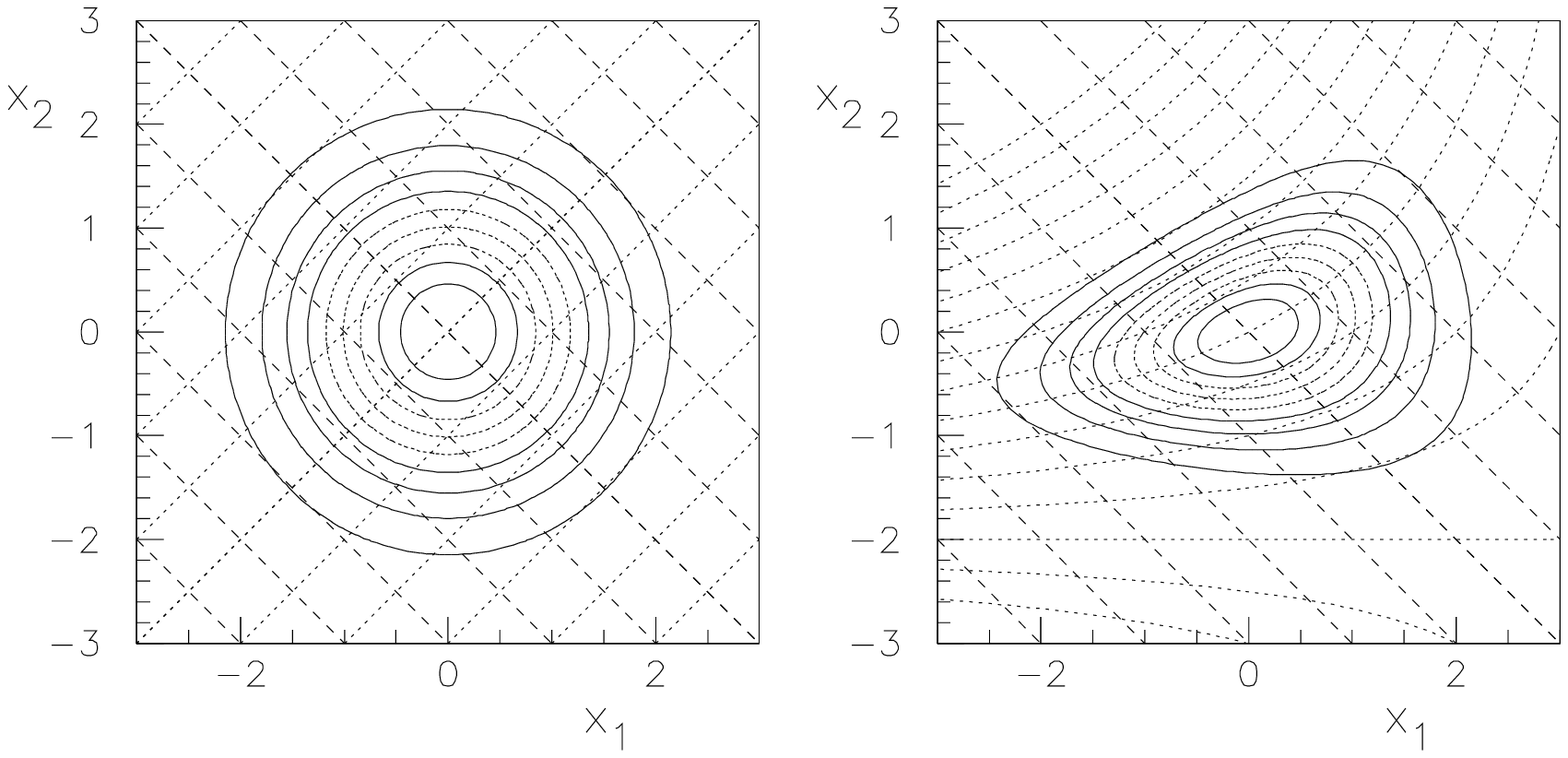}
\centerline {Figure 1: 2-D likelihood functions with lines of
constant $u$ and constant $v$}
}

\vskip \baselineskip

Figure 1 gives an illustration.  The left hand plot shows the
standard double Gaussian (shown as a linear function 
rather than the logarithm, for presentational reasons) as 
a function of $x_1$ and $x_2$. 
The lines of constant $u=x_1+x_2$ run diagonally, from top left to bottom right, and
the lines of constant $v=x_1-x_2$ are orthogonal to them, 
running from bottom left to top right.
For any chosen value of $v$, the profile of the 
likelihood as a function of $u$ is the same 
Gaussian shape, from which 68\% limits can be read off,
the same in each case.  There is a line of constant $v=0$ 
running through the maximum, which follows the
maximum for any chosen $u$.

The right hand plot shows a more interesting  function.
The lines of constant $u=x_1+x_2$ are as before.
The lines of constant $v$ are such that the likelihood
as a function of $u$ along them is the same, up to a 
constant factor. There is a line of constant $v$ through the 
maximum which follows the maximum for any chosen $u$.

This construction shows the limits of the technique.  For 
some given $u$ we plot $L$ as a function of $x_1-x_2$
and compare it with the same curve for $u=0$.  
Then we map the values of $x_1-x_2$ onto the corresponding values at $u=0$ 
at which the log likelihood falls off from the peak by the same amount, and these
give the lines of constant $v$.
If both curves are single peaks then this is readily done and the
mapping is continuous.  If there are multiple peaks then this continuous
mapping is not possible. Thus for a simple peak 
the technique will work, but not if there are secondary peaks
or valleys.   

This generalises readily to the case of several variables. 
The profile
likelihood is a function $\hat L(u)$ where $u=\sum x_i$ and 
$\hat L$ is the maximum value of the likelihood in the $u=constant$ hyperplane.
  
\section Parametrisation of the likelihood function

Thus both questions can be answered if the likelihood 
functions are known.
In general they are not: a quoted result will only give the value and
the positive and negative error. We therefore need a way to reconstruct,
as best we can, the log likelihood function from them, using 
a parametrised curve.

This curve must go through the three points, having
a maximum at the middle one. 
This gives four equations, and hence the curve will
have four parameters, obtainable
from the quoted values of the peak and the positive and negative errors. (The
fourth parameter is an additive constant which controls the value of the
function at its maximum, which is in fact irrelevant for our purposes.) 
It must also behave in a `reasonable' fashion elsewhere. 

Various possibilities have been tried, and tested against the log likelihood
curves where the true value is known, such as the Poisson
and the log of a Gaussian variable.  For simplicity in what follows
we take the quoted value as zero, and work with just $\sigma_+$ and $\sigma_-$
as input parameters.

\subsection Form 1: a cubic 

Adding a cubic term is the obvious step
$$f(x)=-\half(\alpha x^2 + \beta x^3) \eqno(4)$$
with the coefficients readily obtained as 
$\alpha={\sm^3 + \sp^3 \over \sigma_+^2 \sigma_-^2 (\sm+\sp)}$
$\beta={\sm^2 - \sp^2 \over \sigma_+^2 \sigma_-^2 (\sm+\sp)}$.
Extension to several values has some consistency, as adding cubics 
will give another cubic, but associativity is not guaranteed.

This gives curves which will behave
sensibly in the $[x-\sigma^-,x+\sigma^+]$ range, but outside that the 
$x^3$ term produces an unwanted turning point and the curve does not go
to $-\infty$ for large positive and negative $x$. 

\subsection Form 2: A constrained quartic

 A quartic curve can be
constrained to give only one maximum by making the second derivative
a perfect square: 
$$f''(x) = -\half (\alpha  + \beta x)^2 \qquad f(x)=-\half \left({\alpha^2 x^2 \over 2}
+ {\alpha \beta x^3 \over 3 } + {\beta^2 x^4 \over 12}\right) \eqno(5)$$

The parameters are given by
$$\beta={1 \over \sp \sm} \sqrt{6(\sm+\sp)^2 \pm 12 
\sqrt{4 \sp \sm^3 + 4 \sm \sp^3 - 2\sm^4 -2\sp^4} \over
3 \sm^2 + 2 \sm \sp + 3 \sp^2}\eqno(6)$$
Here the negative sign in the expression for $\beta$ should be chosen to give a quartic term 
which is small. In very asymmetric cases 
($\sm$ and $\sp$ differing by more than about  a factor of 2) 
the inner square root
is negative, indicating that there is no solution of the desired form.

Then one solves for $\alpha$ 
$$\alpha=(-) {\beta \sigma \over 3} \pm {\sqrt{36 - 2 \beta^2 \sigma^4} \over 6 \sigma} \eqno(7)$$
for both $\sigma=\sp$ and $\sigma=\sm$, where the $(-)$ minus sign is used
for the $\sm$ case, and selects the solution which
is common to both.

Combination again gives closure, in that the sum of two quartics (with second derivative everywhere negative) is a quartic (with second derivative
everywhere negative.)

This form gives rather better large $x$ behaviour but
is not always satisfactory in the range between $\sm$ and $\sp$.  

\subsection Form 3: Logarithmic

One can also use a
logarithimc approximation
$$f(x) = -\half \left({log(1+\gamma x)\over log \beta}\right)^2\eqno(8)$$
where 
$$\beta=\sigma^+/\sigma^- \qquad \gamma={\sigma_+-\sigma_-\over \sigma_+ \sigma_-}\eqno(9)$$
This is easy to write down and work with, and has some motivation, as
it describes the  expansion/contraction of the abscissa
variable at a constant rate. Its unpleasant features are that it
is undefined  for values of $x$ beyond some point in the 
direction of the smaller error, as $1+\gamma x$ goes negative, and
that it does not give a
parabola in the $\sigma_+=\sigma_-$ limit.

\subsection Form 4: Generalised Poisson

Starting from the Posson likelihood $L(x) = -x + N \ln x - \ln N!$
one can generalise to 
$$f(x) =-\alpha(x+\beta) + \nu  \ln{\alpha(x+\beta)} + const \eqno(10) $$
using  $\nu$,  a continuous variable,  to give skew to the function, and
then scaling and shifting using $\alpha$ and $\beta$.
Putting the maximum at the right place requires $\nu = \alpha \beta$
and thus, adjusting the constant for convenience to make the peak value zero:
$$f(x) =-\alpha x  + \nu  \ln{(1 +{\alpha x \over \nu})} \eqno(10a) $$

Writing $\gamma = \alpha / \nu$ the  equations at $\sm$ and $\sp$ lead to
$${1 - \gamma \sm \over 1 + \gamma \sp}=exp^{-\gamma(\sm+\sp)}\eqno(11)$$
This has to be solved numerically. It has a solution between
$\gamma=0$ and $\gamma=1/\sm$ which can be found by bifurcation. (Attempts to use
more sophisticated algorithms failed.)

Given the value of $\gamma$, $\nu$ is then found from
$$\nu={1 \over 2(\gamma \sp - \ln(1+\gamma \sp))}\eqno(12)$$

This form did fairly
well with many of the tests, but the extraction of the function parameters from
$\sm$ and $\sp$ is inelegantly numerical.

\subsection Form 5: Variable Gaussian (1)

Another function is motivated by 
the Bartlett technique for maximum likelihood
errors [2,5]. This assumes (and indeed justifies)
that the likelihood function for a 
result $\hat x$ from a true value $x$ is described with good accuracy
by a Gaussian whose width depends on the
value of $x$. 
$$ln L(\hat x;x)=- \half \left( {\hat x -x \over \sigma(x)}
\right)^2 \eqno(13)$$
This does not include the $-ln\, \sigma(x)$ term from the denominator
of the Gaussian.  However it turns out [2] that omitting this term
actually improves the accuracy of the 
$\Delta \ln L=-\half$
errors, bringing them into line with the Bartlett form.

We make the further assumption that in the neighbourhood
of interest this variation
in  standard devation is linear
$$\sigma(x)=\sigma+\sigma'(x-\hat x)\eqno(14)$$

$$ln L(\hat x;x)=
 - \half \left({\hat x - x \over \sigma+
\sigma'(x-\hat x)}\right)^2\eqno(15)$$
 the requirement that this go through the $-\half$ points gives
$$\sigma={2 \sp \sm \over \sp + \sm} \qquad 
\sigma'={ \sp -\sm \over \sp + \sm}\eqno(16)  $$
Thus the parameters are easy to find, and when $\sm=\sp$
the symmetric case
is smoothly incorporated.

\subsection Form 6: Variable Gaussian (2)

Still using the Bartlett-inspired form, we could alternatively
take the variance  as linear
$$V(x)=V+V'(x-\hat x)\eqno(17)$$
and 
$$ln L(\hat x;x)=
 - \half {(\hat x - x)^2 \over V+V'(x-\hat x)}\eqno(18)$$
and the parameters are again easy to find, and sensible if $\sm=\sp$ 
$$V = \sm \sp \qquad 
V'= \sp -\sm \eqno(19). $$

\subsection Example: Approximating a Poisson likelihood

\vbox{
\vskip 14.5 cm
\includegraphics{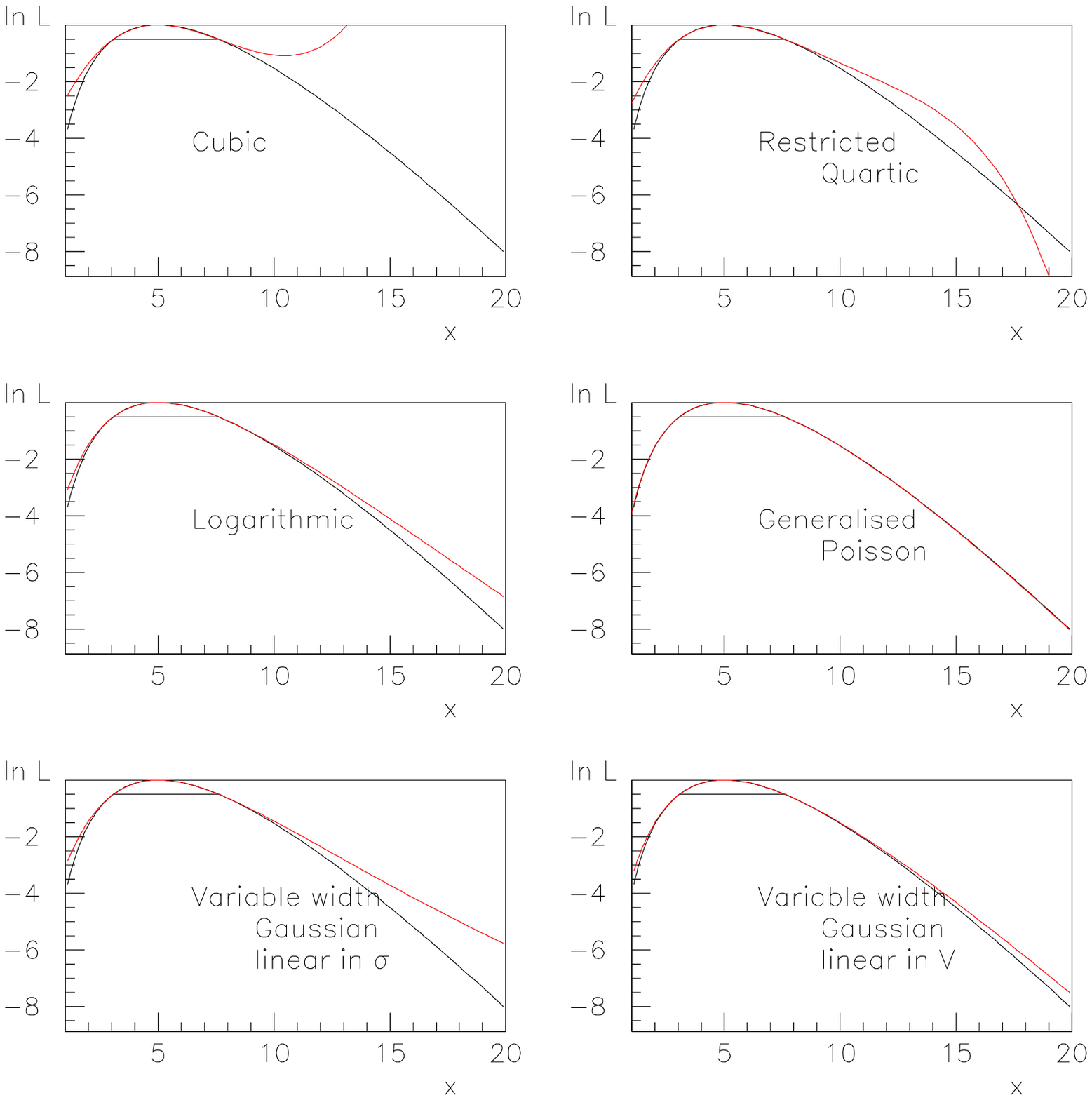}
\centerline {Figure 2: Approximations to a Poisson likelihood}
}
\vfill

Figure 2 shows in black the likelihood function for  Poisson measurement 
of 5 events. 
In red are the approximations, constrained to peak at $x=5$ 
and to go through the 
$-\half $ points, 
indicated by the horizontal line.
They all do well interpolationg in that region, but outside it 
their behavour is very different. The polynomial forms 
diverge significantly from the truth. The logarithmic form does
fairly well, and the generalised Poisson does perfectly (as it should for 
a Poisson likelihood).  
The variable width Gaussian models both do quite well, but the one with
linear variance does noticably better than the form 
linear in the standard deviation

\subsection Example: Approximating a Logarithmic measurement.

\vbox{
\vskip 14.5 cm
\includegraphics{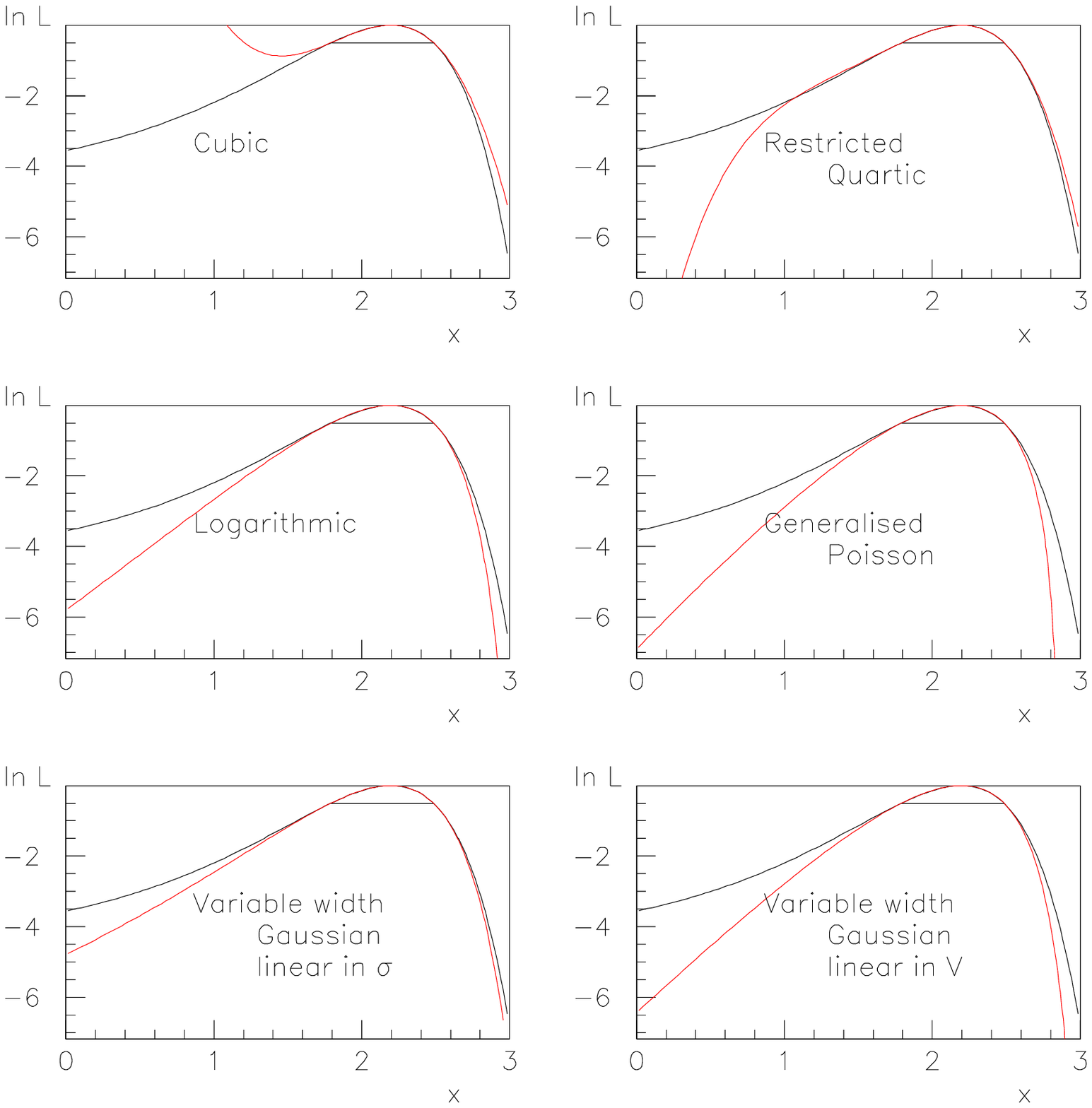}
\centerline {Figure 3: Approximations to the likelihood of the 
log of a Gaussian measuremnet}
}
\vfill
Figure 3 shows the same approximations, fitting a measurement of $x=\ln y$,
where $y$ is a Gaussian measurement with the value $8\pm 3$.

Again, all perform well in the central region, and the
polynomial forms diverge badly outside that region, though the
quartic does adequately on the positive side and
down to about $-2 \sm$ from the peak. The logarithmic curve
does fairly well, but
the generalised Poisson is not so good.
The variable width Gaussians both do well, but in this case
the linear $\sigma$ form does
markedly better than the linear variance form.

We can conclude that the variable width Gaussians are the best
approximation for our purpose, having good descriptive power together
with parameters that are readily 
obtained from Equations 16 or 19, but that 
the choice between the linear $\sigma$ or linear $V$ 
form is one that the user has to make on a case by case basis.
Likelihood functions based on a Poisson measurement will 
be better represented by the linear $V$ form. 

\section Procedure for combination of results

Working with a variable-width Gaussian parametrisation
the likelihood function for a set of measurements $x_i$ is
$$ln L = - \half \sum \left( {\hat x - x_i \over \sigma_i(\hat x)} \right)^2.
\eqno(20)$$

For the linear $\sigma$ form, the position of the maximum is given by the equation
$$ \hat x \sum_i w_i = \sum_i x_i w_i 
\qquad \hbox{\rm with } \qquad
w_i = {\sigma_i \over \left( \sigma_i + \sigma^\prime_i(\hat x - x_i)\right)^3}.
\eqno(21)$$
For the linear $V$ form the corresponding equation is
$$ \hat x \sum_i w_i =\sum_i w_i (x_i-{V_i'\over 2V_i}(\hat x-x_i)^2)
\qquad \hbox{\rm with } \qquad
 w_i={V_i \over (V_i+V_i^\prime (\hat x -x_i))^2}.\eqno(22)$$

The algebra is simple, and has been implemented in 
a Java applet, obtainable under {\tt http://www.slac.stanford.edu/$\sim$barlow/statistics.html}.

Equations 21 and 22 are  nonlinear for $\hat x$, 
and the solution is
found by iteration: ${1 \over N} \sum_i x_i$ is taken 
as a first guess for $\hat x$, and this is used in the right hand side of 
the equation
to give an improved value.
The implementation deems it to have converged if the step size
is less that $10^{-6}$ of the total range of interest, defined as from
$- 3 \sm$ below the lowest point to $+3 \sp$ above the highest.
In practice such convergence occurs after a few iterations.

The $\Delta log L=-\half$ points of the function of Equation 20 
are also found numerically. The function
is reasonably linear over the region where the iteration is performed, and again
convergence is rapid: an initial value is taken, inspired by
Equation (1), as the inverse root sum of the inverse squares of the 
positive or negative, as appropriate, errors. A small step is taken, until
the $-\half$ line is crossed, and successive linear interpolation
is then done until the value is within $10^{-7}$ of 0.5.
Again, only a few iterations are required for a typical case.

The value of the function at the peak gives the $\chi^2$ for the 
result, and this can be used to judge the
compatibility of the different results. (The number of degrees of freedom is
just one less that the number of values being combined.)

\vbox{
\vskip 11 cm
\includegraphics{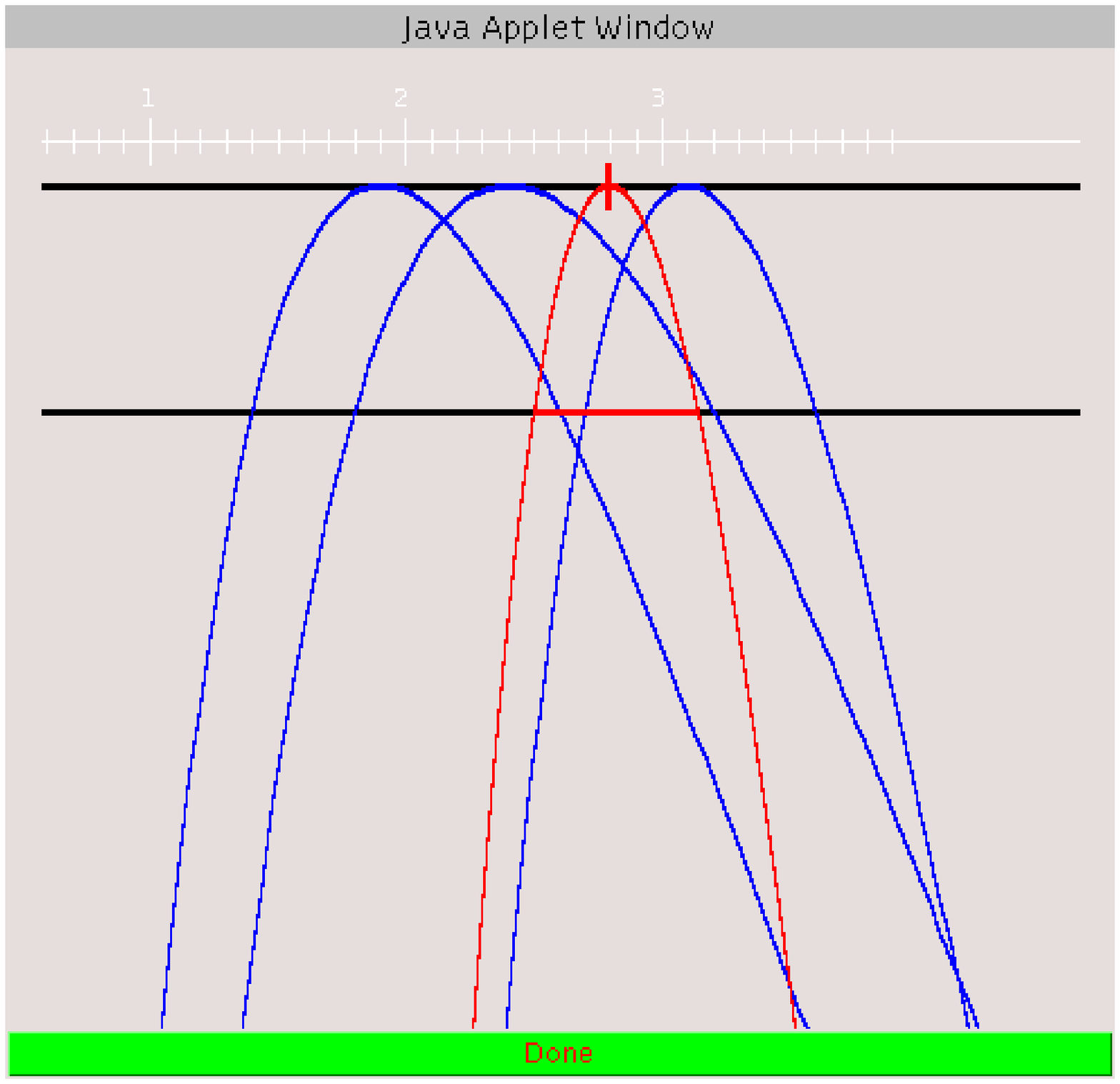}
\centerline {Figure 4:  Three parametrised likelihood curves and their sum
}
}
Figure 4 shows the graphical result of combining
$1.9^{+0.7}_{-0.5}$ with $2.4^{+0.6}_{-0.8}$ and $3.1_{-0.4}^{+0.5}$.
The upper black line shows the peak value (which, as mentioned earlier, is not
relevant and therefore set to zero). The lower black line shows $\ln L=-\half$
The 3 blue curves are the three parametrised likelihood curves (using linear 
$\sigma$). It can be seen that they do indeed each go through their
3 known values correctly. Otherwise we have no precise
knowledge of what they should
look like, but they are apparently well behaved.

The red curve is the sum of the three blue curves (again, adjusted to have a peak
value of zero.)  The position of the peak, found as described above, is
indicated by the short vertical red line, and the horizontal red line
indicates the 68\% confidence interval, again obtained as described above.
One can thus verify by eye that the numerical techniques are giving 
sensible answers.

Results are also given numerically, as shown in Figure 5. Values and errors
are given, and each measurement may be specified as being linear
in $\sigma$ or $V$ using the right hand button.  On pressing the bottom
left button, the graph above is drawn and the numerical values displayed.
There are also facilities to add more values (up to a limit of 10).

\vbox{
\vskip 5 cm
\includegraphics{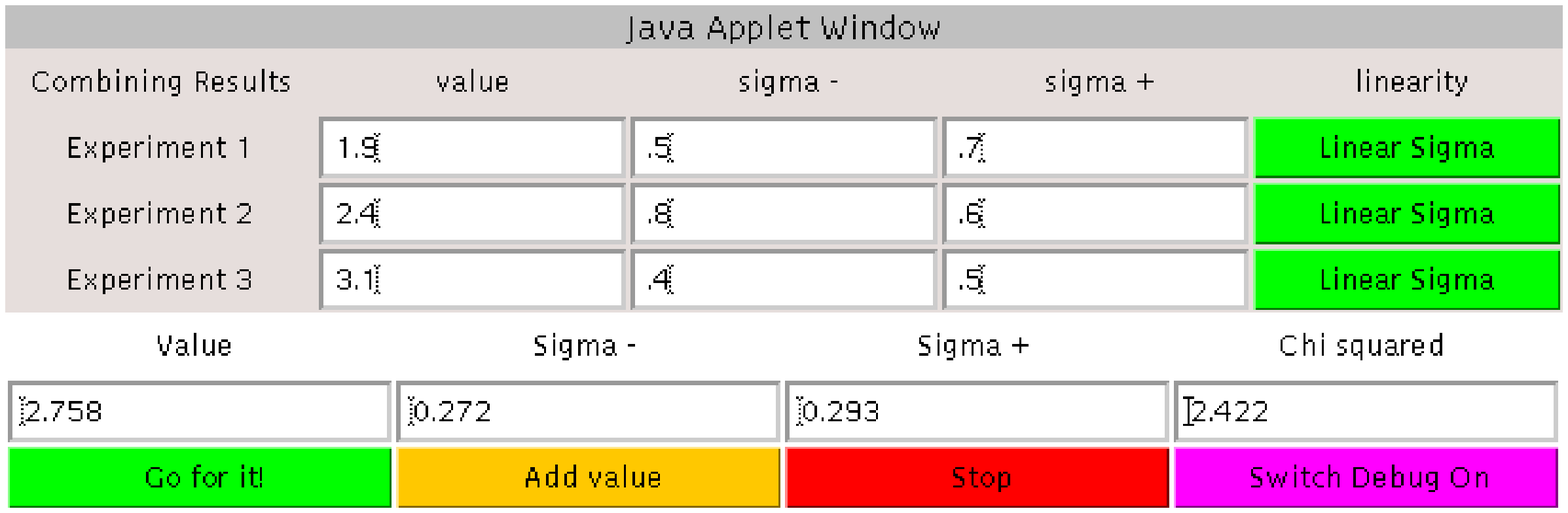}
\centerline {Figure 5: The user interface, showing input values,
output values and options }
}

\subsection Example of combination of results

Suppose a counting experiment sees 5 events. The result is
quoted (using the $\Delta \ln L=-\half$ errors, even though
this is a case where the full Neyman errors could be given) as 
$\AE 5 1.916 2.581 $.  Suppose further that it is repeated and the
same result is obtained.
With the knowledge of the details we can obtain the combined result just by halving the
total measurement of $10_{-2.838}^{+3.504}$ to give an exact
answer of $5_{-1.419}^{+1.752}$. But in general we would not know this
and just be given the measurements, and combine them using the
above method. 
This (using the linear variance model)
gives a combined result of $5_{-1.415}^{+1.747}$.
So the combined result is exact, with
discrepancies only in the fourth decimal place of the errors.

Table 1 shows these, together with the values obtained from other
pairs of results with the same sum.

\vskip \baselineskip

\vbox{
\baselineskip 16 pt
\hbox{\hskip 3 cm \vbox{  \halign{$#$ & $#$ &\qquad $ #$ &\qquad $ #$  \cr
x_1&x_2&\hbox{\rm Linear $\sigma$} &\hbox{\rm Linear $V$}
\cr
\noalign{\hrule}\cr
5_{-1.916}^{+2.581} & 5_{-1.916}^{+2.581} & 5.000_{-1.408}^{+1.737}& 5.000_{-1.415}^{+1.747}\cr
\AE 6 2.128 2.794 & \AE 4 1.682 2.346 & \AE 5.000 1.432 1.778 & \AE 5.000 1.425 1.758  \cr
\AE 7 2.323 2.989  & \AE 3 1.416 2.080 & \AE 5.038 1.529 1.936 & \AE 5.009 1.456 1.793 \cr
\AE 8 2.505 3.171 & \AE 2 1.102 1.765 & \AE 5.402 1.826 2.368 & \AE 5.055 1.515 1.855 \cr
\AE 9 2.676 3.342 & \AE 1 0.6983 1.358 &\AE 7.350 2.548 3.149 & \AE 5.203 1.605 1.942 \cr
}}}
\centerline{Table 1: Combining results in a case of two
samples from the same Poisson distribution}
}
\vskip \baselineskip

This shows that the technique, especially
with the linear variance model, works very
well.  There are discrepancies, but these are reasonable given
the assumptions that have had to be made.  It is worth pointing out that the
larger discrepancies of the final two rows are produced by rather unlikely
experimental circumstances - the probability of 10 events being split 9:1
or even 8:2 between the two experimental runs is small. (This shows
up in their $\chi^2$ values which are large enough to flag a warning.) 

\vfill\eject

\section Procedure for Combination of Errors

To combine errors when
the likelihoods are not given in full, and only the errors are
available, we again parameterise them by the variable Gaussian model
$$ln L(\vec x)
=-\half \sum_i \left( {x_i \over \sigma_i+\sigma^\prime_i x_i}\right)^2
\hbox{\rm  or }  { x_i^2 \over V_i+ V^\prime_i x_i}
\eqno(23)$$
where the $x_i$ represent deviations from the quoted result. Their
total is $u=\sum_i x_i$ and 
to find $\hat L(u)$ the sum of Equation 23 is maximised, subject to the 
constraint
$\sum x_i = u$.
The method of undetermined multipliers gives the solution as
$$x_i = u {w_i \over \sum_j w_j}
\eqno(24)$$

$$\qquad \hbox{\rm where } w_i = {( \sigma_i+\sigma^\prime_i x_i)^3 \over 2 \sigma_i}
\qquad \hbox{\rm or } \qquad {\left( V_i+V'_i x_i \right)^2\over 2 V_i + V'_i x_i}\eqno(25)$$

This is an non-linear set of equations. However 
a solution can be mapped out, starting
at $u=0$ for which all the $x_i$ are zero. Increasing $u$ in small 
amounts, Equation 24 is used to give the small the changes in the $x_i$, and the weights are then re-evaluated using Equation 25.

This has also been implemented by a Java program obtainable at the
web address mentioned above.  It has a similar user interface panel, and
displays the form of $\hat L(u)$ used to read off the total
$\Delta \ln L=-\half$ errors.

\subsection An example of combination of errors

Suppose that $N$  events have been observed in an experiment,
and to extract the 
signal the number of background events must be subtracted.
We suppose that there are several such sources, determined by separate
experiments, and that, for simplicity,  these do not have to be scaled; 
the backgrounds were
determined by running the apparatus, in the absence of 
signal, for the same period of time as the actual experiment.

Suppose that two backgrounds are measured, one giving 4 events and the other 5.
These are reported as 
$\AE 4 1.682 2.346 $ and $\AE 5 1.916 2.581 $.
(again using the $\Delta ln L=-\half$ errors.)
 This method
gives the combined error as ${\ }^{+3.333}_{-2.668}$. 
However in this case where the backgrounds are combined with equal weight,
one could just quote the the total number of background events
as $\AE 9  2.676 3.342 $. The method's error values are
in impressive agreement with this. 
Further examples are given in table 2

\vskip \baselineskip

\vbox{
\hbox {\hskip 2 cm \vbox{
\halign{$#$ \hfil & \ $#$ & \ $# $ & \qquad $#$ & $#$\cr
&{\rm Linear}&\sigma&{\rm Linear} & $V$\cr
{\rm Inputs} & \sigma_- & \sigma_+ & \sm & \sp \cr
4+5 &2.653 & 3.310 &  2.668 & 3.333 \cr
3+6 &2.653 & 3.310 &  2.668 & 3.333 \cr
2+7 &2.653 & 3.310 &  2.668 & 3.333 \cr
1+8 &2.654 & 3.313 &  2.668 & 3.333 \cr
3+3+3 &2.630 & 3.278 &  2.659 & 3.323 \cr
1+1+1+1+1+1+1+1+1 & 2.500 & 3.098 & 2.610 & 3.270 
\cr
}}}
\vskip \baselineskip
\centerline{Table 2: Various combinations of Poisson errors which should give
$\sm= 2.676$, $\sp=3.342$}
}
\vskip \baselineskip

\section Conclusions

If the full likelihood functions are not given, then 
there is no exact method for combination
of errors and results with asymmetric statistical errors.
However the procedures decribed here, which work by 
making an approximation to the likelihood function on the basis of 
the quoted value and errors, appear to be reasonably accurate and
robust. They are also easy to implement and user.

\vskip \baselineskip

\leftline{\bf Acknowledgements}

The author gratefully acknowledges the support of the Fulbright Foundation

\vskip \baselineskip

\parindent 0 pt

\leftline{\bf References}

[1] R.J. Barlow:{\it Asymmetric Systematic Errors},
 arXiv physics/0306168, (2003)

[2] R.J. Barlow:{\it A Note on $\Delta ln L =-\half$ errors},
 arXiv physics/0403046, (2004)

[3] M. Schmelling:{\it Averaging Measurements with Hidden Correlations and Asymmetric Errors},
arXiv:hep-ex/0006004, (2000) 

[4] N.Read and D.A.S. Fraser, {\it Likelihood inference in the 
presence of Nuisance Parameters},  Proc. PHYSTAT2003, Ed. L.Lyons, R. Mount, R. Reitmeyer, 
SLAC-PUB R 603 eConf 030908.

[5] M.S. Bartlett: {\it On the Statistical Estimation of Mean Lifetimes,}
Phil. Mag. {\bf 44} 244 (1953),\hfill\break
--- \qquad  {\it Estimation of Mean Lifetimes from Multiple Plate Cloud Chamber Tracks,}
Phil. Mag. {\bf 44} 1407 (1953)

\bye